# Gelation in Multiple Link Systems

Kazumi Suematsu
Institute of Mathematical Science
Ohkadai 2-31-9, Yokkaichi, Mie 512-1216, JAPAN
Fax: +81 (0) 593 26 8052, E-mail address: suematsu@m3.cty-net.ne.jp

## Summary

Within the framework of the random distribution assumption of cyclic bonds, the theory of gelation is extended to mixing systems of the R-$A_g$ + R-$B_{f-g}$ model. The ring concentration of this system has the form:

$$[\Gamma]_{C\to\infty} = \sum_{j=1}^{\infty} \varphi_j \frac{1}{2j}\left\{\left(\tfrac{1}{2}t - \sqrt{s+\left(\tfrac{1}{2}t\right)^2}\right)^j + \left(\tfrac{1}{2}t + \sqrt{s+\left(\tfrac{1}{2}t\right)^2}\right)^j\right\}D_A^j,$$

where the subscript $C$ represents the initial monomer concentration, $J$ the number of functional units needed to form a junction point, $s = (J-1)(\langle g_w\rangle - 1)(\langle(f-g)_w\rangle - 1)/\kappa$ and $t = \left(\tfrac{J}{2}-1\right)\left\{(\langle g_w\rangle - 1) + (\langle(f-g)_w\rangle - 1)/\kappa\right\}$. As soon as the gel point is passed, the concentration of cyclic species should diverge and hence

$$\left(\tfrac{1}{2}t + \sqrt{s+\left(\tfrac{1}{2}t\right)^2}\right)D_A = 1.$$

We know, on the other hand, that the random distribution assumption yields $D_c = D_{co} + p_R$. From these informations, we obtain immediately the gel point expression:

$$D_c = \frac{\sqrt{t^2 + 4s} - t}{2s}\left\{\frac{1 - \mathcal{F}\sum_j(1-1/j)\varphi_j\{\cdots\}\gamma_f}{1 - \mathcal{F}\sum_j \varphi_j\{\cdots\}\gamma_f}\right\},$$

with $\{\cdots\} = \left\{\left(\left[\tfrac{1}{2}t - \sqrt{s+\left(\tfrac{1}{2}t\right)^2}\right]\Big/\left[\tfrac{1}{2}t + \sqrt{s+\left(\tfrac{1}{2}t\right)^2}\right]\right)^j + 1\right\}$ and $\mathcal{F} = \frac{(1+\kappa)J}{4(J-1)}\left(\tfrac{1}{2}t + \sqrt{s+\left(\tfrac{1}{2}t\right)^2}\right)$.

## Key Words

## §1. Introduction

In this report, we derive the mathematical expression of the gel point in multifunctional systems. We give mathematical proofs to the basic equations employed in the theory of gelation. We focus our attention to multiple link systems of the R-$A_g$ + R-$B_{f-g}$ model, which is expected to have wider application such as micell formations in biological systems. In common with other papers of this series [1], we push forward our discussion on the basis of the following three principles and one assumption:

(1) The gel point is divided into the two terms

$$D_c = D(inter) + D(ring). \tag{1}$$

(2) The total ring concentration, $[\Gamma]$, is independent of the initial monomer concentration, $C$, and is a function of $D$ (the extent of reaction) alone.
(3) Branched molecules behave ideally at $C = \infty$.
(4) Assumption I: Cyclic bonds distribute randomly over all bonds.

Introducing Assumption I is important, since it reduces an otherwise intrinsically insoluble problem of polymer physics to an elementary mathematical exercise, leading us to the simple relations: $D(inter) = D_{co}$ and $D(ring) = p_R$, namely,

$$D_c = D_{co} + p_R. \tag{2}$$

In eq. (2), $D_{co}$ denotes the Flory's classic gel point [2] and $p_R$ the fraction of cyclic bonds to the total possible bonds. Since $p_R$ is a function of $D_c$, eq. (2) is an implicit function of $D_c$. Now the gel point problem is analytic and yields neat solutions. The physical meaning of eq. (2) is very clear: Cyclic bonds simply waste functional units (FU), making no contribution to the growth of molecules, so the gel point is exactly equal to the point where the fraction of intermolecular bonds attains the ideal gel point, $D_{co}$. By eq. (2), the problem of finding the gel point reduces to the problem of expressing $p_R$ in terms of $D_c$.

## §2. THEORETICAL

### 2-1. DERIVATION OF EQ. (2)

Consider a mixing system comprised of two different type of monomer units, $\{g_i M_{A_i}\}$ and $\{(f-g)_j M_{B_j}\}$, where $M_{A_i}$ and $M_{B_j}$ are the mole numbers of the A and B type monomers, respectively, and $g_i$ and $(f-g)_j$ are the corresponding functionalities. Let $J$ be the number of FU's to form a junction point on which the two types of the FU's are situated alternately. By the nature of the $R-A_g + R-B_{f-g}$ model, $J$ must be an even integer. It then follows that unit reaction occurs through the merger of $J/2$ A FU's and $J/2$ B FU's, and $J-1$ branches arise. The familiar bimolecular reaction corresponds to the special case of $J=2$.

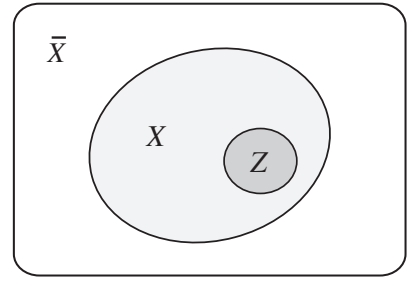

Fig. 1. Representation of $P(Z \cap X)$: $X$={all bonds}, $Z$={cyclic bonds}.

We seek the probability, $\alpha$, that a given FU on a R-$A_g$ branching unit leads, via R-$B_{f-g}$ units, to the next R-$A_g$ unit. This is an application of the first-success problem in statistics, the general permutation being $AA-(BB)_n-AA$ ($n = 0 \sim \infty$). Let $P(Z|X)$ be a conditional probability defined by $P(Z|X) = P(Z \cap X)/P(X)$, where $X$ is a set of all bonds and $Z$ is a set of all cyclic bonds, so that $P(Z|X) = p_R/D_A$ (Fig. 1). It is clear that

for $n = 0$:
$$\alpha_0 = D_A(J/2-1)(1-P(Z|X))(\langle g_w \rangle - 1);$$

for $n = 1$:
$$\alpha_1 = D_A(J/2)(1-P(Z|X))(\langle (f-g)_w \rangle - 1) D_B(J/2)(1-P(Z|X))(\langle g_w \rangle - 1);$$

for $n = 2$:
$$\alpha_2 = D_A(J/2)(1-P(Z|X))(\langle (f-g)_w \rangle - 1) D_B \cdot \mathscr{J} \cdot (J/2)(1-P(Z|X))(\langle g_w \rangle - 1);$$

...
...

by induction,
for $n = \ell$:
$$\alpha_\ell = D_A(J/2)(1-P(Z|X))(\langle (f-g)_w \rangle - 1) D_B \cdot \mathscr{J}^{\ell-1} \cdot (J/2)(1-P(Z|X))(\langle g_w \rangle - 1), \tag{3}$$

where $\mathscr{J} = (J/2-1)(1-P(Z|X))(\langle (f-g)_w \rangle - 1) D_B$. The probability, $\alpha$, is obtained by summing up all these terms; i.e., $\alpha = \sum_{\ell=0}^{\infty} \alpha_\ell$. When $\alpha$ exceeds the critical value, $\alpha = 1$, an infinite network should arise, so the gel point is

$$\alpha_c = D_A \left\{ \left(\tfrac{J}{2}-1\right)(1-P(Z|X))(\langle g_w \rangle - 1) + \frac{\left(\tfrac{J}{2}\right)^2 (1-P(Z|X))^2 (\langle g_w \rangle - 1)(\langle (f-g)_w \rangle - 1) D_B}{1 - \left(\tfrac{J}{2}-1\right)(1-P(Z|X))(\langle (f-g)_w \rangle - 1) D_B} \right\} = 1. \tag{4}$$

From Fig. 1, it is seen that $P(Z|X) = p_R/D_A$. Eq. (4) then gives

$$s(D_c - p_R)^2 + t(D_c - p_R) - 1 = 0, \tag{5}$$

with

$$s = (J-1)(\langle g_w \rangle - 1)(\langle (f-g)_w \rangle - 1)/\kappa,$$

$$t = \left(\tfrac{J}{2}-1\right)\left\{(\langle g_w \rangle - 1) + (\langle (f-g)_w \rangle - 1)/\kappa\right\},$$

and $\kappa = \sum_j (f-g)_j M_{B_j} / \sum_i g_i M_{A_i}$ the relative molar concentration of B FU to A FU. Solving eq. (5), we arrive at the relationship of eq. (2):

$$D_c = \frac{\sqrt{t^2 + 4s} - t}{2s} + p_R, \tag{6}$$

The first term of eq. (6) represents the ideal gel point. For $J = 2$, we recover the known result [1]:

$$D_c = 1 \Big/ \sqrt{(\langle g_w \rangle - 1)(\langle (f-g)_w \rangle - 1)/\kappa} + p_R. \tag{7}$$

## 2-2. CLUSTER PROFILE IN R-A$_G$ + R-B$_{F-G}$ MODEL

We seek the total number of FU's in the $n$th generation on a mean cluster without rings. First we consider the homogeneous mixture of the R-A$_g$ + R-B$_{f-g}$ model. Consider an $m$-tree which has $m$ unreacted A-Type FU's in the root (1st generation). Let $N(F)_n$ be the number of FU's in the $n$th generation. It is clear that

$$N(F)_1 = g - m;$$

$$N(F)_2 = (g-m)\left\{\left(\tfrac{J}{2}-1\right)(g-1) + \tfrac{J}{2}(f-g-1)\right\};$$

$$N(F)_3 = (g-m)\left\{\left(\tfrac{J}{2}-1\right)(g-1)D_A\left[\left(\tfrac{J}{2}-1\right)(g-1) + \tfrac{J}{2}(f-g-1)\right]\right.$$

$$\left. + \tfrac{J}{2}(f-g-1)D_B\left[\tfrac{J}{2}(g-1) + \left(\tfrac{J}{2}-1\right)(f-g-1)\right]\right\};$$

...
...

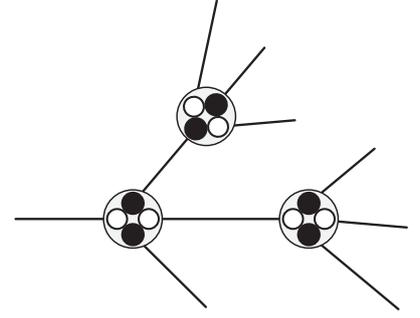

Fig. 2. Representation of the AA-BB multiple branching. (○): AA unit; (●): BB unit. Large circles show junction points.

In general we can write

$$N(F)_n = (g-m)\left[a_n(g-1) + b_n(f-g-1)\right]; \tag{8}$$

$$N(F)_{n+1} = (g-m)\left\{a_n(g-1)D_A\left[\left(\tfrac{J}{2}-1\right)(g-1) + \tfrac{J}{2}(f-g-1)\right]\right.$$

$$\left. + b_n(f-g-1)D_B\left[\tfrac{J}{2}(g-1) + \left(\tfrac{J}{2}-1\right)(f-g-1)\right]\right\},$$

or

$$N(F)_{n+1} = (g-m)\left[a_{n+1}(g-1) + b_{n+1}(f-g-1)\right], \tag{9}$$

with

$$a_{n+1} = \left(\tfrac{J}{2}-1\right)(g-1)D_A \cdot a_n + \tfrac{J}{2}(f-g-1)D_B \cdot b_n;$$

$$b_{n+1} = \tfrac{J}{2}(g-1)D_A \cdot a_n + \left(\tfrac{J}{2}-1\right)(f-g-1)D_B \cdot b_n. \tag{10}$$

Eq. (10) satisfies the matrices:

$$\begin{pmatrix} a_{n+1} \\ b_{n+1} \end{pmatrix} = \begin{pmatrix} \epsilon_{11} & \epsilon_{12} \\ \epsilon_{21} & \epsilon_{22} \end{pmatrix} \begin{pmatrix} a_n \\ b_n \end{pmatrix}. \tag{11}$$

Here

$$\epsilon_{11} = \left(\tfrac{J}{2}-1\right)(g-1)D_A; \quad \epsilon_{12} = \tfrac{J}{2}(f-g-1)D_B;$$

$$\epsilon_{21} = \tfrac{J}{2}(g-1)D_A; \quad \epsilon_{22} = \left(\tfrac{J}{2}-1\right)(f-g-1)D_B \tag{12}$$

and

$$(a_2, b_2) = \left(\tfrac{J}{2}-1, \tfrac{J}{2}\right);$$

$$(a_3, b_3) = \left(\left(\tfrac{J}{2}-1\right)^2(g-1)D_A + \left(\tfrac{J}{2}\right)^2(f-g-1)D_B, \tfrac{J}{2}\left(\tfrac{J}{2}-1\right)\left[(g-1)D_A + (f-g-1)D_B\right]\right).$$

$$\tag{13}$$

Let us write eq. (11) in the form:

$$\tilde{N}_{n+1} = \tilde{A} \cdot \tilde{N}_n. \tag{14}$$

To find the solution, transform eq. (14) as

$$\tilde{N}_n = \tilde{A}^{n-2} \cdot \tilde{N}_2. \tag{15}$$

Applying the Cayley-Hamilton theorem to eq. (15), one has

$$\begin{pmatrix} a_n \\ b_n \end{pmatrix} = \begin{pmatrix} \frac{\beta^{n-2}-\alpha^{n-2}}{\beta-\alpha}\epsilon_{11} + \frac{\alpha^{n-2}\beta-\alpha\beta^{n-2}}{\beta-\alpha} & \frac{\beta^{n-2}-\alpha^{n-2}}{\beta-\alpha}\epsilon_{12} \\ \frac{\beta^{n-2}-\alpha^{n-2}}{\beta-\alpha}\epsilon_{21} & \frac{\beta^{n-2}-\alpha^{n-2}}{\beta-\alpha}\epsilon_{22} + \frac{\alpha^{n-2}\beta-\alpha\beta^{n-2}}{\beta-\alpha} \end{pmatrix} \begin{pmatrix} J/2-1 \\ J/2 \end{pmatrix},$$

which yields

$$\begin{pmatrix} a_n \\ b_n \end{pmatrix} = \begin{pmatrix} \left(\frac{J}{2}-1\right)\left[\frac{\beta^{n-2}-\alpha^{n-2}}{\beta-\alpha}\epsilon_{11} + \frac{\alpha^{n-2}\beta-\alpha\beta^{n-2}}{\beta-\alpha}\right] + \frac{J}{2}\frac{\beta^{n-2}-\alpha^{n-2}}{\beta-\alpha}\epsilon_{12} \\ \left(\frac{J}{2}-1\right)\frac{\beta^{n-2}-\alpha^{n-2}}{\beta-\alpha}\epsilon_{21} + \frac{J}{2}\left[\frac{\beta^{n-2}-\alpha^{n-2}}{\beta-\alpha}\epsilon_{22} + \frac{\alpha^{n-2}\beta-\alpha\beta^{n-2}}{\beta-\alpha}\right] \end{pmatrix}, \quad (16)$$

where $\alpha$ and $\beta$ are the eigenvalues of the determinant: $\det(\tilde{A}-\lambda\tilde{E})=0$, and $\alpha \geq \beta$. By eq. (8), the total number of FU's of the tree, therefore, can be expressed in the form:

$$N(F) = \sum_{k=1}^{\infty} N(F)_k = \frac{C_1}{1-\alpha} + \frac{C_2}{1-\beta}. \quad (17)$$

Clearly an infinite gel must appear at $\alpha = 1$ or $\beta = 1$, but it can be proven rigorously that $\beta$ is a negative quantity $(\beta \leq 0)$ for all $J \geq 1$, so that the gelation should occur at

$$\alpha = \tfrac{1}{2}\left[\epsilon_{11}+\epsilon_{22}+\sqrt{(\epsilon_{11}+\epsilon_{22})^2 - 4(\epsilon_{11}\epsilon_{22}-\epsilon_{12}\epsilon_{21})}\right] = 1, \quad (18)$$

which gives $\epsilon_{11}+\epsilon_{22}-\epsilon_{11}\epsilon_{22}+\epsilon_{12}\epsilon_{21}=1$. With the help of eq. (12) and $D_A/D_B = \kappa$, one has

$$(J-1)[(g-1)(f-g-1)/\kappa]D_c^2 + (\tfrac{J}{2}-1)[(g-1)+(f-g-1)/\kappa]D_c - 1 = 0, \quad (19)$$

with $D_c$ being the critical extent of reaction of A FU's. Eq. (19) is the critical condition for the ideal tree model with no rings, and corresponds to the $P(Z|X)=0$ (no rings) case in the previous report. As expected, for $J=2$ eq. (19) reduces to the familiar result:

$$[(g-1)(f-g-1)/\kappa]D_c^2 = 1. \quad (20)$$

The extension of the above discussion to the multifunctional system of $\{g_i M_{A_i}\}$ and $\{(f-g)_j M_{B_j}\}$ is easy; i.e., it can be accomplished simply by the transformation:

$$g - m \rightarrow g_i - m; \quad g \rightarrow \langle g_w \rangle; \quad f - g \rightarrow \langle (f-g)_w \rangle. \quad (21)$$

The extension to the system with ring formation is also straightforward. Note that the $D_c$ in eq. (19) represents the ideal gel point, while by eq. (2), $D_{co} = D_c - p_R$. Thus substituting this into eq. (19), one has

$$s(D_c - p_R)^2 + t(D_c - p_R) - 1 = 0. \quad (22)$$

Eq. (20) is just eq. (5) in the preceding section, with $s$ and $t$ having the same meanings as given in eq. (5).

## 2-3. ESTIMATION OF RING CONCENTRATION

To derive the concentration of cyclic species in polymer solutions [3], we consider the transition probability in the unit reaction. Let $p(ring\ j)$ be the probability that one functional unit (FU) on a $j$-chain jumps to form a $j$-ring, and $p(inter)$ the probability that the FU jumps to form a intermolecular bond, where the $j$-chain and the $j$-ring denote a chain and a ring comprised of $j$ repeating units, respectively. Since the reaction must be either the ring formation or the intermolecular reaction, we must have

$$\sum_{j=1}^{\infty} p(ring\ j) + p(inter) = 1. \quad (23)$$

These quantities can be expressed, using the rate equations, in the forms:

$$\sum_{j=1}^{\infty} p(ring\ j) = \sum_{j=1}^{\infty} \frac{v_{R_j}}{v_L + v_R}; \quad (24)$$

$$p(inter) = \frac{v_L}{v_L + v_R}, \quad (25)$$

where $v_L$ denotes the velocity of the intermolecular reaction and $v_{R_j}$ that of the $j$-ring formation, and $v_R = \sum_j^{\infty} v_{R_j}$, as defined earlier. $p(ring\ j)$ can be equated with the number fraction of a $j$-ring to be formed in the unit reaction, $\delta u$, and is generally a function of $D$. The total number of $j$-rings to be formed is, therefore, calculated by taking the summation:

$$N_{R_j} = \int p(ring\ j) \cdot \delta u. \quad (26)$$

We derive the expression of the ring concentration in the asymptotic limit of $C \rightarrow \infty$, where the relative frequency of cyclization to the intermolecular reaction is negligible. Experiments have shown that $v_L \gg v_R$ for sufficient high concentration, so that the approximation

$$p(ring\ j) = v_{R_j}/(v_L + v_R) \doteq v_{R_j}/v_L, \quad (27)$$

is valid. Substituting eq. (27) into eq. (26), one has

$$N_{R_j} \doteq \int \left(v_{R_j}/v_L\right) \cdot \delta u. \tag{28}$$

As discussed earlier, as $C \to \infty$, the excluded volume effects are expected to vanish, then we can make use of the ideal behavior of branched molecules with no rings and no excluded volume effects (the tree approximation).

Consider a general case in which the reaction proceeds by way of the merger of $J$ FU's. The familiar polymerization reaction is, thus, a special case of $J = 2$. Take notice of A-type FU's and we have the equality, $\delta D_A = (J/2)\delta u / \sum_i g_i M_{A_i}$, because for every unit reaction, $J/2$ A FU's are consumed. Thus

$$N_{R_j} = \frac{2\sum_i g_i M_{A_i}}{J} \int \left(v_{R_j}/v_L\right) \cdot \delta D_A. \tag{29}$$

This formula is frequently used as a basic equation to derive closed solutions for specific models.

In order to express $v_{R_j}$ with experimentally measurable quantities, we must evaluate the total number, $\phi_j$, of chances of $j$-ring formation. Let $\mathscr{P}$ be the probability that one end of a $j$-chain enters the small volume $v$ around the other end. To avoid complication, we solve a special case in which an A-A chain and a B-B chain have an equal length and backbone so that one can make use of the probability, $\mathscr{P}$, common to all $j$-chains with different permutations of, say, AA-AA-AA, AA-BB-AA and so forth. Let $\nu$ be the number of cyclic bonds. Clearly the velocity of the $j$-ring formation is of the form:

$$v_{R_j} = d\nu_j/dt \propto \mathscr{P}\phi_j + \mathcal{O}, \tag{30}$$

where $\mathcal{O}$ represents the higher order cyclization (dual, triple, ...) which becomes less probable as $C \to \infty$. To carry out the rigorous calculation of $v_{R_j}$, $\phi_j$ must be evaluated for each chain species, namely, $\phi_j = \phi_{j,AA} + \phi_{j,AB} + \phi_{j,BB}$. According to eqs. (8) and (9), it follows that $N(A)_j = (g_i - m)a_j(\langle g_w \rangle - 1)$. The total number of AA chains is then

$$\phi_{j,AA} = \tfrac{1}{2}\sum_i M_{A_i} \sum_{m=0}^{g_i} m \cdot \binom{g_i}{m}(1-D_A)^m D_A^{g_i-m}(g_i-m)a_j(\langle g_w \rangle - 1)(1-D_A), \tag{31}$$

so that

$$\phi_{j,AA} = \tfrac{1}{2}\sum_i M_{A_i} g_i(g_i-1)a_j(\langle g_w \rangle - 1)D_A(1-D_A)^2. \tag{32}$$

By the symmetry of AA and BB, we have for $\phi_{j,BB}$

$$\phi_{j,BB} = \tfrac{1}{2}\sum_k M_{B_k}(f-g)_k\left((f-g)_k - 1\right)b'_j\left(\langle (f-g)_w \rangle - 1\right)D_B(1-D_B)^2, \tag{33}$$

but $b'_j$ is one of the solutions of the form:

$$\begin{pmatrix} a'_n \\ b'_n \end{pmatrix} = \begin{pmatrix} \frac{J}{2}\left[\frac{\beta^{n-2}-\alpha^{n-2}}{\beta-\alpha}\epsilon_{11} + \frac{\alpha^{n-2}\beta-\alpha\beta^{n-2}}{\beta-\alpha}\right] + \left(\frac{J}{2}-1\right)\frac{\beta^{n-2}-\alpha^{n-2}}{\beta-\alpha}\epsilon_{12} \\ \frac{J}{2}\frac{\beta^{n-2}-\alpha^{n-2}}{\beta-\alpha}\epsilon_{21} + \left(\frac{J}{2}-1\right)\left[\frac{\beta^{n-2}-\alpha^{n-2}}{\beta-\alpha}\epsilon_{22} + \frac{\alpha^{n-2}\beta-\alpha\beta^{n-2}}{\beta-\alpha}\right] \end{pmatrix}, \tag{34}$$

slightly different from eq. (16).

Since $N(B)_j = (g_i - m)b_j(\langle (f-g)_w \rangle - 1)$, we have for $\phi_{j,AB}$

$$\phi_{j,AB} = \sum_i M_{A_i} \sum_{m=0}^{g_i} m \cdot \binom{g_i}{m}(1-D_A)^m D_A^{g_i-m}(g_i-m)b_j\left(\langle (f-g)_w \rangle - 1\right)(1-D_B),$$

so that

$$\phi_{j,AB} = \sum_i M_{A_i} g_i(g_i-1)b_j\left(\langle (f-g)_w \rangle - 1\right)D_A(1-D_A)(1-D_B). \tag{35}$$

Now the unit reaction occurs between $J/2$ A FU's and $J/2$ B's. There are, in general, $\binom{N}{J}$ ways to choose $J$ units from $N$ units. If $N$ is a large number and $N \gg J$, then it follows that $\binom{N}{J} \cong N^J/J!$. And thus

$$dv_{jAA}/dt = I\mathscr{P}\phi_{j,AA}\frac{\left\{\sum_i g_i M_{A_i}(1-D_A)\right\}^{\frac{J}{2}-2}}{\left(\frac{J}{2}-2\right)!}\frac{\left\{\sum_j (f-g)_j M_{B_j}(1-D_B)\right\}^{\frac{J}{2}}}{\left(\frac{J}{2}\right)!}(v/V)^{J-2};$$

$$dv_{jAB}/dt = I\mathscr{P}\phi_{j,AB}\frac{\left\{\sum_i g_i M_{A_i}(1-D_A)\right\}^{\frac{J}{2}-1}}{\left(\frac{J}{2}-1\right)!}\frac{\left\{\sum_j (f-g)_j M_{B_j}(1-D_B)\right\}^{\frac{J}{2}-1}}{\left(\frac{J}{2}-1\right)!}(v/V)^{J-2};$$

$$dv_{j,BB}/dt = I\mathcal{P}\phi_{j,BB} \frac{\left\{\sum_i g_i M_{A_i}(1-D_A)\right\}^{\frac{J}{2}}}{\left(\frac{J}{2}\right)!} \frac{\left\{\sum_j (f-g)_j M_{B_j}(1-D_B)\right\}^{\frac{J}{2}-2}}{\left(\frac{J}{2}-2\right)!} (v/V)^{J-2},$$

where $I$ is a constant and $v/V$ represents the probability that a FU enters a small volume, $v$, around another FU. The $v_L$, on the other hand, has the form:

$$v_L = I \cdot \frac{\left\{\sum_i g_i M_{A_i}(1-D_A)\right\}^{J/2} \left\{\sum_j (f-g)_j M_{B_j}(1-D_B)\right\}^{J/2}}{(J/2)!\,(J/2)!} (v/V)^{J-1}. \tag{36}$$

It is clear that one can put $a_j = \mu_{a_j} D_A^{j-2}$, $b_j = \mu_{b_j} D_A^{j-2}$ and $b'_j = \mu_{b'_j} D_A^{j-2}$ with $\mu$ being a constant independent of $D$. Since $p(ring\ j) = [d(v_{j,AA} + v_{j,AB} + v_{j,BB})/dt]/v_L$, substituting these relations into eq. (29) and putting $[\Gamma] = \sum_{j=1}^{\infty} N_{R_j}/V$, one obtains

$$[\Gamma]_{C\to\infty} = \sum_{j=1}^{\infty} \varphi_j \frac{1}{2j} \left\{ \left(\tfrac{1}{2}t - \sqrt{s + \left(\tfrac{1}{2}t\right)^2}\right)^j + \left(\tfrac{1}{2}t + \sqrt{s + \left(\tfrac{1}{2}t\right)^2}\right)^j \right\} D_A^j, \tag{37}$$

where $\varphi_j = \mathcal{P}/v$ is the relative cyclization frequency introduced earlier [1], [3].

One can check the soundness of eq. (37). As soon as the gel point is passed, the production of rings is expected to diverge. Thus, taking account of the boundary condition, $0 \le D_A \le 1$, one has at $D = D_c$

$$\left(\tfrac{1}{2}t + \sqrt{s + \left(\tfrac{1}{2}t\right)^2}\right) D_A = 1, \tag{38}$$

which giving

$$D_c = \frac{1}{2s}\left(\sqrt{t^2 + 4s} - t\right), \tag{39}$$

in agreement with the result mentioned in eq. (6). When $J = 2$, $t = 0$ so that the inner term $\{\cdots\}$ in eq. (37) becomes

$$\{\cdots\} = \left\{\left(-\sqrt{(\langle g_w\rangle - 1)(\langle (f-g)_w\rangle - 1)/\kappa}\right)^j + \left(\sqrt{(\langle g_w\rangle - 1)(\langle (f-g)_w\rangle - 1)/\kappa}\right)^j\right\}. \tag{40}$$

All odd powers in eq. (40) should cancel out. Then altering the index of the relative cyclization frequency as $\varphi_{2k} \to \varphi_k$, one quickly recovers the previous result [3]:

$$[\Gamma]_{C\to\infty} = \sum_{j=1}^{\infty} \varphi_j \frac{1}{2j}\left[(\langle g_w\rangle - 1)(\langle (f-g)_w\rangle - 1)D_A^2/\kappa\right]^j. \tag{41}$$

It will be useful to examine the behavior of the inner term in eq. (37). To do so, let $\varsigma(j) = \tfrac{1}{2}\{\cdots\}D_A^j$. In Fig. 3, $\varsigma(j)$ is plotted as a function of $D_A$ and $j$ (generation) for the special case of $J = 4$, $\langle g_w\rangle = 2$, $\langle (f-g)_w\rangle = 3$, and $\kappa = 1$. As one can see, $\varsigma(j)$ is finite for all $j$'s for $D \le D_c = \tfrac{1}{12}(-3 + \sqrt{33})$, but diverges strongly for $D > D_c$. Important is the fact that eq. (37) is convergent at $D = D_c$.

The result of Fig. 3 is easily comprehensible from the mathematical form of $\varsigma(j)$. At $D = D_c$, it follows that

$$\varsigma(j) = \frac{1}{2}\left\{\left(\frac{\tfrac{1}{2}t - \sqrt{s + \left(\tfrac{1}{2}t\right)^2}}{\tfrac{1}{2}t + \sqrt{s + \left(\tfrac{1}{2}t\right)^2}}\right)^j + 1\right\}. \tag{42}$$

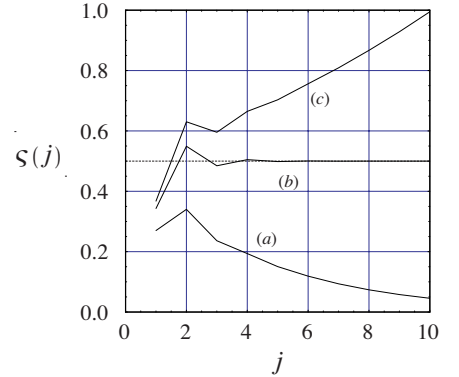

Fig. 3. Plot of $\varsigma(j)$ as a function of $D_A$ and $j$ (generation) for $J = 4$, $\langle g_w\rangle = 2$, $\langle (f-g)_w\rangle = 3$ and $\kappa = 1$. (a): $D_A = 0.18$; (b): $D_A = D_c$; (c): $D_A = 0.245$.

The first term of eq. (42) satisfies $|(\cdots)^j| \le 1$. For small $j$'s, it fluctuates around 0 changing the sign, and as $j \to \infty$, it rapidly approaches 0, so that $\varsigma(j) \to 1/2$, in accord with the behavior shown in Fig. 3 (curve (b)).

## 2-4. Calculation of $p_R$

The fraction, $p_R$, of cyclic bonds to total possible bonds is defined by

$$p_R = \frac{total\ number\ of\ cyclic\ bonds}{total\ number\ of\ possible\ bonds}. \tag{43}$$

To relate eq. (43) with experimentally measurable quantities, we consider the simplest case of $J = 3$ of the R-$A_f$ model. In Fig. 4, an example of $f = 2$ and $J = 3$ is shown; larger (filled) circles represent monomer units and open circles FU's, where merger takes place among three FU's to create one junction point ($\otimes$). Let the statement of this process be $p \Rightarrow q$, where $p$ represents "the merger of the three FU's" and $q$ "the occurrence of $k$ bonds". Our question is, "What is the number of $k$ in this unit reaction?" To answer this question, it is more convenient to ask the contrapositive, $\bar{q} \Rightarrow \bar{p}$, logically equivalent; namely, the question, "How many bond should be broken to split the resultant *3-mer* into the original monomer units?" As is seen from Fig. 4, there are 3!/2! ways to recover the original state; each requires exactly two bond-breaks. Thus the answer is $k = 2$.

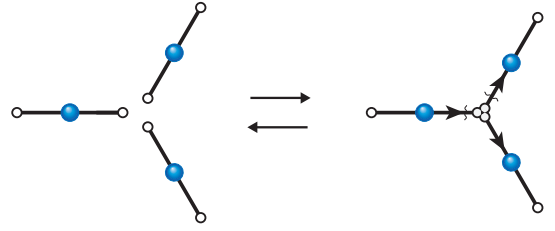

Fig. 4. An example of unit reaction of the R-$A_f$ model. Open circles denote functional units and filles circles branching units. The unit reaction takes place by the merger of three functional units.

In general, $J - 1$ bonds arise as a result of the merger of $J$ FU's. Since there are $C_f$ FU's in the system, the total number of possible bonds is $(J-1/J)C_f$. Recall that every ring has only one cyclic bond. Then the fraction of cyclic bonds to all possible bonds can be expressed as

$$p_R = \frac{J[\Gamma]}{(J-1)C_f}. \tag{44}$$

For $J = 2$, eq. (44) reduces to the known result [1], as expected.

The result can be easily extended to the R-$A_g$ + R-$B_{f-g}$ model, where $J - 1$ bonds arise by the merger of $J/2$ A FU's and $J/2$ B FU's. Thus

$$p_R = \frac{J[\Gamma]}{2(J-1)C_{f,A}} = (1+\kappa)\frac{J[\Gamma]}{2(J-1)}\gamma_f. \tag{45}$$

with $C_{f,A} = \sum_i g_i M_{A_i}/V$ representing the concentration of A FU's, $\gamma_f$ the reciprocal of the total FU concentration defined by $\gamma_f = 1/C_f = V/\left(\sum_i g_i M_{A_i} + \sum_j (f-g)_j M_{B_j}\right)$, and $\kappa$ the relative mole number of B FU's to A FU's as defined in Section 2-1. Combining eq. (6) with eq. (45), we obtain

$$D_c = \frac{\sqrt{t^2+4s}-t}{2s} + (1+\kappa)\frac{J[\Gamma]}{2(J-1)}\gamma_f. \tag{46}$$

## 2-5. Derivation of Gel Point

Only task remained is to unify eqs. (37) and (46). We note, however, that eq. (37) unfortunately breaks down as soon as the ideal gel point is exceeded; i.e., beyond the gel point, the concentration of rings diverges strongly (see Fig. 3). To resolve this problem, we make use of the linear approximation of eq. (26). Recall that at high monomer concentration, the concentration of rings is independent of the monomer concentration itself. Thus we can approximate $[\Gamma]$ as a function of $D$ alone: At $D = D_c$, we can write

$$[\Gamma] = \mathcal{C}(D_c) = \frac{2C_{f,A}}{J}\sum_{j=1}^{\infty}\int_0^{D_c}\frac{(v_{R_j}/v_L)}{1+(v_R/v_L)}dD_A. \tag{47}$$

Experiments have shown that $[\Gamma]$ is a continuous and monotonic function of $D_A$ in the interval, $D_{co} \leq D_A \leq D_c$. Now expand eq. (47) with respect to $D_c = D_{co}$ to yield

$$\mathcal{C}(D_c) = \mathcal{C}(D_{co}) + \frac{\mathcal{C}'(D_{co})}{1!}(D_c - D_{co}) + \cdots. \tag{48}$$

If the system under consideration is in sufficient high concentration so that $[\Gamma]$ is a function of $D$ alone, we may use eq. (37) in place of eq. (47). Then from eq. (48), one has

$$[\Gamma] \cong \sum_{j=1}^{\infty}\varphi_j\{\cdots\}/2j + \frac{1}{2D_{co}}\sum_{j=1}^{\infty}\varphi_j\{\cdots\}(D_c - D_{co}), \tag{49}$$

where

$$\{\cdots\} = \left\{\left(\frac{\frac{1}{2}t - \sqrt{s+\left(\frac{1}{2}t\right)^2}}{\frac{1}{2}t + \sqrt{s+\left(\frac{1}{2}t\right)^2}}\right)^j + 1\right\}. \tag{50}$$

In eq. (50), we have made use of the classical relation (39). Substituting eq. (49) into eq. (46), we arrive at finally

$$D_c = \frac{\sqrt{t^2 + 4s} - t}{2s}\left\{\frac{1 - \mathcal{F}\sum_j (1 - 1/j)\varphi_j\{\cdots\}\gamma_f}{1 - \mathcal{F}\sum_j \varphi_j\{\cdots\}\gamma_f}\right\}, \qquad (51)$$

where

$$\mathcal{F} = \frac{(1+\kappa)J}{4(J-1)}\left(\tfrac{1}{2}t + \sqrt{s + \left(\tfrac{1}{2}t\right)^2}\right). \qquad (52)$$

In Fig. 5, $D_c$ is plotted as a function of $\gamma_f$ and $J$ [4], [5]. In common with all other models, the gel point shifts upward with increasing dilution ($\gamma_f$). This is because of the increasing frequency of cyclization with dilution [5]. Since $D_c$ is restrained by the boundary condition, $0 \leq D(inter) + D(ring) \leq 1$, there is a critical regime of dilution beyond which $D(inter)$ can not reach the ideal gel point. This point is the critical dilution, $\gamma_{f,c}$, the intersections of the $D_c = 1$ line and the $D_c - \gamma_f$ curves [1]. As one can see from Fig. 5, the critical dilution occurs abruptly, in direct contrast to the sudden appearance of a macroscopic gel molecule.

## §3. Conclusion

We have derived the closed solution of the concentration of cyclic species in the multifunctional system (eq. (37)). The result is equal to the formula derived in the preceding paper. In the present work, the mathematical expression of the inner term $\{\cdots\}$ has been greatly simplified. Eq. (37) is a generalization of the solutions for the $A_g$-R-$B_{f-g}$ and R-$A_g$ + R-$B_{f-g}$ models.

Our question central to the works of this series is whether Assumption I is correct; i.e., whether or not, eq. (2) is mathematically exact. Within the framework of the principle of equireactivity under the athermal condition, there is no way to distinguish cyclic bonds from intermolecular bonds: Once a cyclic bond is formed, all memories of intramolecular bonding are lost. In this sense, the assumption that cyclic bonds distribute randomly over all bonds seems mathematically sound. Only way to settle this problem, however, depends on the extensive examination of the theory by observations.

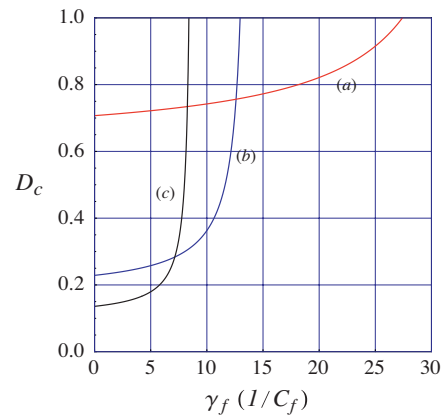

Fig. 5. Plot of $D_c$ as a function of $\gamma_f$ and $J$ for $\langle g_w \rangle = 2, \langle (f-g)_w \rangle = 3$ and $\kappa = 1$. (a): $J = 2$; (b): $J = 4$; (c): $J = 6$.

## References

1. (a) Suematsu, K. Phys. Chem. Chem. Phys., 2002, **4**, 4161;
   (b) Suematsu, K. Adv. Polym. Sci., 2002, **156**, 137;
   (c) Suematsu, K. Macromol. Theory Simul., 2003, **12**, 476;
   (d) Suematsu, K. http://arXiv.org/cond-mat/0410137.
2. Flory, P. J. Principles of Polymer Chemistry, Cornell University Press, Ithaca, New York, 1953.
3. (a) Spouge, J. L., Proc. R. Soc. Lond., 1983, A **387**, 351;
   (b) Spouge, J. L., J. Stat. Phys., 1986, **43**, 363.
4. (a) Wile, L. L. Ph.D. dissertation, Columbia University, New York, 1945;
   (b) Gordon, M. and Scantlebury, G. R. J. Chem. Soc., 1967, **B**, 1.
5. (a) Ilavsky, M. and Dusek, K., Macromolecules, 1986, **19**, 2139;
   (b) Ilavsky, M. and Dusek, K., Polymer, 1983, **24**, 981;
   (c) Matejka, L. and Dusek, K., Polymer Bulletin, 1980, **3**, 489.